\documentclass[10pt,letterpaper]{article}
\usepackage[top=0.85in,left=2.75in,footskip=0.75in]{geometry}

\usepackage{amsmath,amssymb}

\usepackage{changepage}

\usepackage[utf8x]{inputenc}

\usepackage{textcomp,marvosym}

\usepackage{cite}

\usepackage{nameref,hyperref}

\usepackage[right]{lineno}

\usepackage{microtype}
\DisableLigatures[f]{encoding = *, family = * }

\usepackage[table]{xcolor}

\usepackage{array}

\usepackage{multirow}

\newcolumntype{+}{!{\vrule width 2pt}}

\newlength\savedwidth

\newcommand\thickhline{\noalign{\global\savedwidth\arrayrulewidth\global\arrayrulewidth 2pt}%
\hline
\noalign{\global\arrayrulewidth\savedwidth}}

\newcommand{\Eref}[1]{Eq.\,(\ref{#1})}
\newcommand{\Fref}[1]{{Fig.\,\ref{#1}}}

\raggedright
\usepackage[aboveskip=1pt,labelfont=bf,labelsep=period,justification=raggedright,singlelinecheck=off]{caption}

\makeatletter
\renewcommand{\@biblabel}[1]{\quad#1.}
\makeatother

\date{}

\usepackage{lastpage,fancyhdr,graphicx}
\usepackage{epstopdf}
\fancyheadoffset[L]{2.25in}
\fancyfootoffset[L]{2.25in}
\begin{document}
\vspace*{0.2in}

\begin{flushleft}
{\Large
\textbf\newline{Estimation of the Mechanical Properties of the Eye through the Study of its Vibrational Modes} % Please use "title case" (capitalize all terms in the title except conjunctions, prepositions, and articles).
}
\newline
\\
M.A. Aloy\textsuperscript{1,*\Yinyang},
J.E. Adsuara\textsuperscript{1\Yinyang},
P. Cerd\'a-Dur\'an\textsuperscript{1\Yinyang},
M. Obergaulinger\textsuperscript{1\Yinyang},
J.J. Esteve-Taboada\textsuperscript{2\Yinyang},
T. Ferrer-Blasco\textsuperscript{2\Yinyang},
R. Mont\'es-Mic\'o\textsuperscript{2\Yinyang}
\\
\bigskip
\textbf{1} Department of Astronomy and Astrophysics. University of Valencia. Spain.
\\
\textbf{2} Department of Optics and Optometry and Vision Sciences. University of Valencia. Spain.
\\
\bigskip

% Insert additional author notes using the symbols described below. Insert symbol callouts after author names as necessary.
% 
% Remove or comment out the author notes below if they aren't used.
%
% Primary Equal Contribution Note
\Yinyang These authors contributed equally to this work.

% Use the asterisk to denote corresponding authorship and provide email address in note below.
* miguel.a.aloy@uv.es

\end{flushleft}
\section*{Abstract}
 Measuring the eye’s mechanical properties in vivo and with minimally
  invasive techniques can be the key for individualized solutions to a
  number of eye pathologies. The development of such techniques
  largely relies on a computational modelling of the eyeball and, it
  optimally requires the synergic interplay between experimentation
  and numerical simulation. In Astrophysics and Geophysics the remote
  measurement of structural properties of the systems of their realm
  is performed on the basis of (helio-)seismic techniques.  As a
  biomechanical system, the eyeball possesses normal vibrational modes
  encompassing rich information about its structure and mechanical
  properties. However, the integral analysis of the eyeball
  vibrational modes has not been performed yet.  Here we show that the
  vibrational eigenfrequencies of the human eye fall in the interval
  100\,Hz -- 10\,MHz.  The eyeball normal modes have frequencies
  separable from the typical range of phasic physiological phenomena
  (e.g., respiration, pulse). We find that compressible vibrational
  modes may release a trace on high frequency changes of the
  intraocular pressure, while incompressible normal modes could be
  registered analyzing the scattering pattern that the motions of the
  vitreous humour leave on the retina. Existing contact lenses with
  embebed devices operating at high sampling frequency could be used
  to register the microfluctuations of the eyeball shape we obtain.
  We advance that an inverse problem to obtain the mechanical
  properties of a given eye (e.g., Young's modulus, Poisson ratio)
  measuring its normal frequencies is doable. These measurements can
  be done using non-invasive techniques, opening very interesting
  perspectives to estimate the mechanical properties of eyes {\em in
    vivo}. Future research might relate various ocular pathologies
  with anomalies in measured vibrational frequencies of the eye.

% Please keep the Author Summary between 150 and 200 words
% Use first person. PLOS ONE authors please skip this step. 
% Author Summary not valid for PLOS ONE submissions.   
\section*{Author Summary}
Inspired by the (helio-)seismic techniques customary employed in
Geophysics and Astrophysics, we build the foundations of a novel set
of techniques for the remote (non-invasive) measurement of the eyeball
biomechanical properties. Employing a simplified model of the human
eye, we obtain that the typical normal mode frequencies of the eye
fall above 100\,Hz. Remarkably, these frequencies are significantly
different from other phasic physiological phenomena and can be very
likely measured with existing devices. The result is rather robust, as
the dependence of the eigenfrequencies on the elastic moduli is very
modest in the typical ranges of variations of these parameters for
human eyes.

%\linenumbers

% Use "Eq" instead of "Equation" for equation citations.
\section*{Introduction}

Obtaining the mechanical properties of the human eye is fundamental
for the future development of artificial materials that can be
employed as substitutes for natural tissues
\cite{Bonfiglio:2015}. Measuring the eye's mechanical properties {\em
  in vivo} and with minimally invasive techniques can be the key for
individualized solutions to a number of eye pathologies.  The
development of such techniques largely relies on a computational
modelling of the eyeball \cite{Uchio:1999} and, it optimally requires
the synergic interplay between experimentation and numerical
simulation \cite{Hugar:2013}.

The eye is a complex organ consisting of several functional and
mutually interacting parts \cite{Ethier:2004}. The most important ones
from the mechanical point of view are the cornea, lens, vitreous,
sclera and retina. Each of these elements holds distinctive mechanical
properties that are closely related to their respective anatomic
functionality. Changes in the mechanical properties may entail a
number of pathologies or even a loss of functionality
\cite{Morita:2012}. Reciprocally, damage inflicted to a healthy eye
may result in changes in its elastic and mechanical properties
\cite{Hirneiss:2011}. 
% The earliest attempts to build a mechanical model of the eye assumed it was spherically symmetric\cite{Weinbaum1965}.
%The purpose of that model was to roughly predict the overall intraocular phenomena of the
 %eye (specially the variations of the intraocular pressure). 
The mechanical modelling of the human eye is a field that has gained
relevance to rationalize the physiology and pathology of the eye
\cite{Hugar:2013}. The field is exponentially developing pace to pace
with our ability of implementing more complex models on modern
computers \cite{Chung:2016}. Our knowledge of the mechanical
properties of the eye has basically come through three different ways:
experimentation, {\em in vivo} monitoring, or computational
modelling. We develop our work in the later framework.

 Measuring the elasticity properties of the different tissues forming
 an eye is challenging. Very often, the determination of mechanical
 properties of the eye results from a mechanical interaction with its
 different parts \cite{Jones:1992,Zauberman:1972,deGuillebon:1972}.
%Work in vitro has shown that the retina possesses elastic properties, where if free retina is pulled, it stretches until it tears\cite{deGuillebon:1972}. 
 In addition to standard mechanical testing, the cornea has been
 characterized through high-resolution microscopy techniques
 \cite{Lombardo:2009}, as well as with the Ocular Response Analyzer
 \cite{Luce:2005,Terai:2012}. Likewise, multiple studies have examined
 the overall biomechanical properties of the sclera
 \cite{Sigal:2005,Sigal:2009}.
 % , including the effective stiffness as well as the thickness and globe size.
Ultrasound biomicroscopy has been used to measure the scleral
 thickness \cite{Lam:2005,Oliveira:2006}.
%, with the limitation that the scleral thickness close to the optical nerve head
% –which is of particular interest in glaucoma– cannot be resolved. 
 Magnetic Resonance Imaging (MRI) techniques applied {\em in vivo}
 resulted inaccurate because of the random eye movement of the
 patients, though it is possible to use MRI scans to produce 3D models
 of the corneoscleral shells in post-mortem patients
 \cite{Norman:2010}.

 Novel non-invasive techniques need to be devised to measure the
 mechanical properties of the eye. Here we show that these properties
 are related to the normal vibrational modes of the eyeball, i.e., to
 the periodic variations of matter inside of the eyeball resulting
 from perturbations with respect to its equilibrium state. We have
 been inspired by the extensive use of the remote measurements of
 normal-mode related physical quantities in Geophysics and Astrophysics. For
 instance, the solar interior is routinely scanned by means of
 helioseismic techniques, which are based on the measurement of the
 global resonant oscillations of the
 Sun \cite{Vorontsov:2002}. Likewise, employing the principles of
 asteroseismology, neutron star interiors are proven \cite{Israel:2005,Strohmayer:2005}
 in an attempt to decipher
 the equation of state for matter at nuclear densities.
We believe that this treatment opens up a new set of techniques for remote measuring of the eyes structural properties.

\section*{Materials and Methods}
\label{sec:methodology}

The oscillations under consideration in our model are free elastic
vibrations, which we assume may arise when applying generic stresses,
e.g., on the sclera or the cornea. We tackle the numerical calculation
of the vibrational eigenfrequencies and eigenmodes of the human eye
under a number of simplifying assumptions. We model the eyeball as a
\emph{spherical}, \emph{homogeneous} and \emph{isotropic}
\emph{elastic} solid ball with axial symmetry. While assuming that the
eyeball is axially symmetric is very well justified, the assumptions
of homogeneity and isotropy are certainly not the most accurate
possible. However, these assumptions serve for the primary purpose of
reducing the dependence of the constitutive equation only to two
elastic constants or moduli of the eye material: the Young's modulus
$E$, and the Poisson ratio $\sigma$. In this simplified framework, we
will compute, first analytically and afterwards numerically, the
eigenfrequencies of the model attempting to grasp the essential
mechanics of an average human eye.

% Results and Discussion can be combined.
\section*{Results}

\begin{figure}[!h]
\includegraphics[width=0.99\columnwidth]{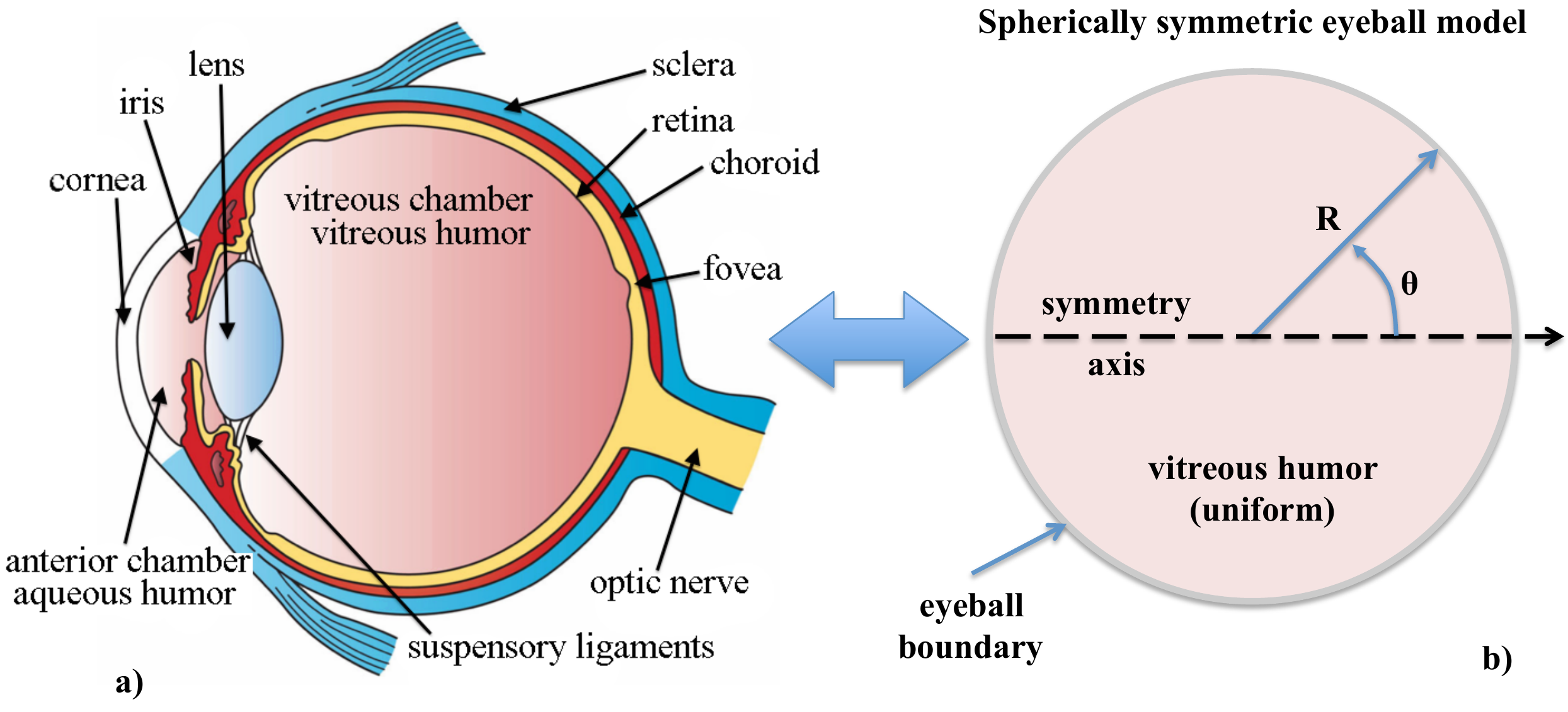}
  \caption{{\bf Simplified mechanical model of the eyeball.} Left:
    transversal cut of the human eye with the different structural
    parts annotated in it (source: Wikipedia). Right: spherically
    symmetric, homogeneous and isotropic eyeball model employed in
    this work.}
  \label{fig:scheme}
\end{figure}

As we have mentioned above, we model the eyeball as a spherically
symmetric, homogeneous and isotropic elastic solid ball
(\Fref{fig:scheme}).  This simplification
allows us to use known analytical solutions (\nameref{sec:analytic}) in
other physics disciplines (e.g., seismology \cite{Visscher:1991} or
gravitational wave physics \cite{Rue:1996,Love:1944}) to calibrate our numerical
code (described in Sect.\,\nameref{sec:code}).
% Place figure captions after the first paragraph in which they are cited.

\subsection*{Numerical code}
\label{sec:code}

Since we aim to employ non-trivial boundary conditions, we are forced
to solve numerically the eigenvalue problem at hand. We have developed
a code that solves the eigenvalue problem set by the Navier-Cauchy
equation discretizing the eyeball sphere on a two-dimensional grid of
nodes in spherical coordinates
($0 \le r \le R,\: 0 \le \theta \le \pi$).  As a first step, we have
assumed the elastic moduli to be uniform throughout the spatial
grid. However, there is no restriction to implement elastic moduli
that depend on the location in the eyeball. This is important because
it may enable us to improve the degree of realism of our model for the
vibrational modes of the eye, in particular, by using different
elastic moduli for the sclera, the cornea, the lens, and the vitreous
humour.

First, we will rewrite \Eref{eq:eigPro} for the numerical part with
the analytical solutions available in the previous section.  In
spherical coordinates, and under the assumption of axisymmetry, i.e.,
neglecting the $\varphi$-dependence, the displacements can be written
as
$$u_i = (u_r(r,\theta), u_{\theta}(r,\theta),u_{\varphi}(r,\theta)),$$ 
and satisfy the following three equations:
\begin{eqnarray}
-\frac{\mu}{\rho} \Delta u_{r} - \frac{\lambda + \mu}{\rho} u_{j,jr} = p^2 u_{r} \label{eq:eigPro_r} \\
-\frac{\mu}{\rho} \Delta u_{\theta} - \frac{\lambda + \mu}{\rho} u_{j,j\theta} = p^2 u_{\theta} \label{eq:eigPro_theta} \\
-\frac{\mu}{\rho} \Delta u_{\varphi} - \frac{\lambda + \mu}{\rho} u_{j,j\varphi} = p^2 u_{\varphi}  \label{eq:eigPro_varphi}.
\end{eqnarray}

Under the assumption of axisymmetry, all the derivatives with respect
$\varphi$ vanish and \Eref{eq:eigPro_varphi} decouples from the other
two Equations (\ref{eq:eigPro_r}) and (\ref{eq:eigPro_theta}).  We
obtain the toroidal modes from the latter equation:
\begin{eqnarray}
-\frac{\mu}{\rho} \Delta u_{\varphi} = p^2 u_{\varphi}.
\label{eq:torEq}
\end{eqnarray}
with traction boundary conditions, $u_{\varphi,r} = 0$. The spheroidal modes result from Equations (\ref{eq:eigPro_r}) and (\ref{eq:eigPro_theta}):
\begin{eqnarray}
-\frac{\mu}{\rho} \Delta u_{r} - \frac{\lambda +
  \mu}{\rho} \partial_{r} \bigg [ \frac{1}{r^2}\partial_{r}r^2 u_r +
  \frac{1}{r\sin\theta} \partial_{\theta} \sin \theta u_{\theta} \bigg ] = p^2 u_{r}  \label{eq:sphEq1} \\
-\frac{\mu}{\rho} \Delta u_{\theta} - \frac{\lambda +
  \mu}{r\rho} \partial_{\theta} \bigg [ \frac{1}{r^2} \partial_{r} r^2
  u_r + \frac{1}{r\sin\theta} \partial_{\theta} \sin\theta u_{\theta} \bigg ]  = p^2 u_{\theta}  \label{eq:sphEq2}
\end{eqnarray}
also with traction boundary conditions, $u_{r,r} = 0; u_{\theta,r} = 0$.  Equations (\ref{eq:sphEq1}) and
(\ref{eq:sphEq2}) can be cast as an eigenvalue equation, $L u = \lambda u$, with the vectorial operator
\begin{equation}
L u := - \frac{\mu}{\rho} \Delta u_i  - \frac{\lambda + \mu}{\rho} u_{j,ji}
\end{equation}
with $i=r,\theta$, i.e., the equations are coupled due to the axisymmetry hypothesis. If we explicitly insert spherical coordinates, then: 
\begin{equation}
L u = \bigg (- \frac{\mu}{\rho} \Delta u_r  - \frac{\lambda + \mu}{\rho} u_{j,jr}, - \frac{\mu}{\rho} \Delta u_\theta  - \frac{\lambda + \mu}{\rho} u_{j,j\theta} \bigg )
\end{equation}
and, for clarity, in matrix form we have: 
\begin{equation}
\left[ \begin{array}{cc}
a_{rr} & a_{r \theta} \\
a_{\theta r} & a_{\theta \theta} \\
\end{array} 
\right ] 
\left[ \begin{array}{c}
u_{r} \\
u_{\theta} \\
\end{array} 
\right ] 
\end{equation}
where expansion of  the operator yields:
\begin{eqnarray}
 a_{rr} &\hspace{-0.2cm}:=\hspace{-0.2cm}& - \frac{\lambda + 2 \mu}{\rho} \partial_{rr} - \frac{2(\lambda + 2 \mu)}{\rho r} \partial_{r} - \frac{\mu}{\rho r^2} \partial_{\theta \theta} - \frac{\mu \cot \theta}{\rho r^2} \partial_{\theta} + \frac{2(\lambda + 2 \mu)}{\rho r^2} \\
 a_{r \theta} &\hspace{-0.2cm}:=\hspace{-0.2cm}&  -\frac{(\lambda +\mu)\cot \theta}{\rho r} \partial_r + \frac{\lambda +3 \mu }{\rho  r^2} \partial_{\theta} -\frac{\lambda +\mu }{\rho  r} \partial_{r \theta} 
+ \frac{\cot \theta (\lambda +3 \mu )}{\rho  r^2}\\
a_{\theta r} &\hspace{-0.2cm}:=\hspace{-0.2cm}& \frac{2 (\lambda +2 \mu )}{\rho  r^2} \partial_{\theta} -\frac{\lambda +\mu }{\rho  r} \partial_{r \theta} \\
a_{\theta \theta} &\hspace{-0.2cm}:=\hspace{-0.2cm}& -\frac{\mu }{\rho } \partial_{rr} -\frac{2 \mu }{\rho  r} \partial_{r} + -\frac{\lambda +2 \mu }{\rho  r^2} \partial_{\theta \theta} -\frac{(\lambda +2 \mu )\cot \theta }{\rho  r^2} \partial_{\theta} 
+ \frac{(\lambda +2 \mu )\csc ^2 \theta}{\rho  r^2}
\end{eqnarray}

In both types of modes, we compute in a first step the eigenvalues
(vibrational frequencies), and in a second step the eigenfunctions
(normal displacements).
For the eigenvalues we simply compute the zeros of the characteristic
polynomial. In practice, working in logarithmic space is advantageous
because it reduces the magnitude of coefficients of the polynomial.
Knowing the family of eigenvalues, we compute the kernel for each one
of them. We substitute each eigenvalue into the corresponding
equation, \Eref{eq:torEq} or
Eqs.\,(\ref{eq:sphEq1})-(\ref{eq:sphEq2}), obtaining an elliptic
equation.  With this, we subtract the eigenvalue from the diagonal of
the matrix produced by the discretization of the elliptic operator,
and proceed further solving the corresponding system of equations by
direct numerical inversion of the matrix of the system.  As the rank
of this matrix cannot be complete, we will obtain the compatible but
indeterminate solution as a function of some of the variables (either
one or two variables for the toroidal and the spheroidal case,
respectively).

\subsubsection*{Code calibration}
\label{sec:calibration}

We calibrate the code by comparing the frequencies computed with our
numerical code and the corresponding analytic values at a density of
$\rho=1\,$kg\,m$^{-3}$, an elastic moduli of $E=2.5$\,Pa,
$\sigma=0.25$ and a radius of the sphere of $R=1\,$m. Note that these
values do not correspond to a typical human eye. They are employed for
numerical convenience.

\begin{figure*}[!h]
\centerline{\includegraphics[width=\columnwidth]{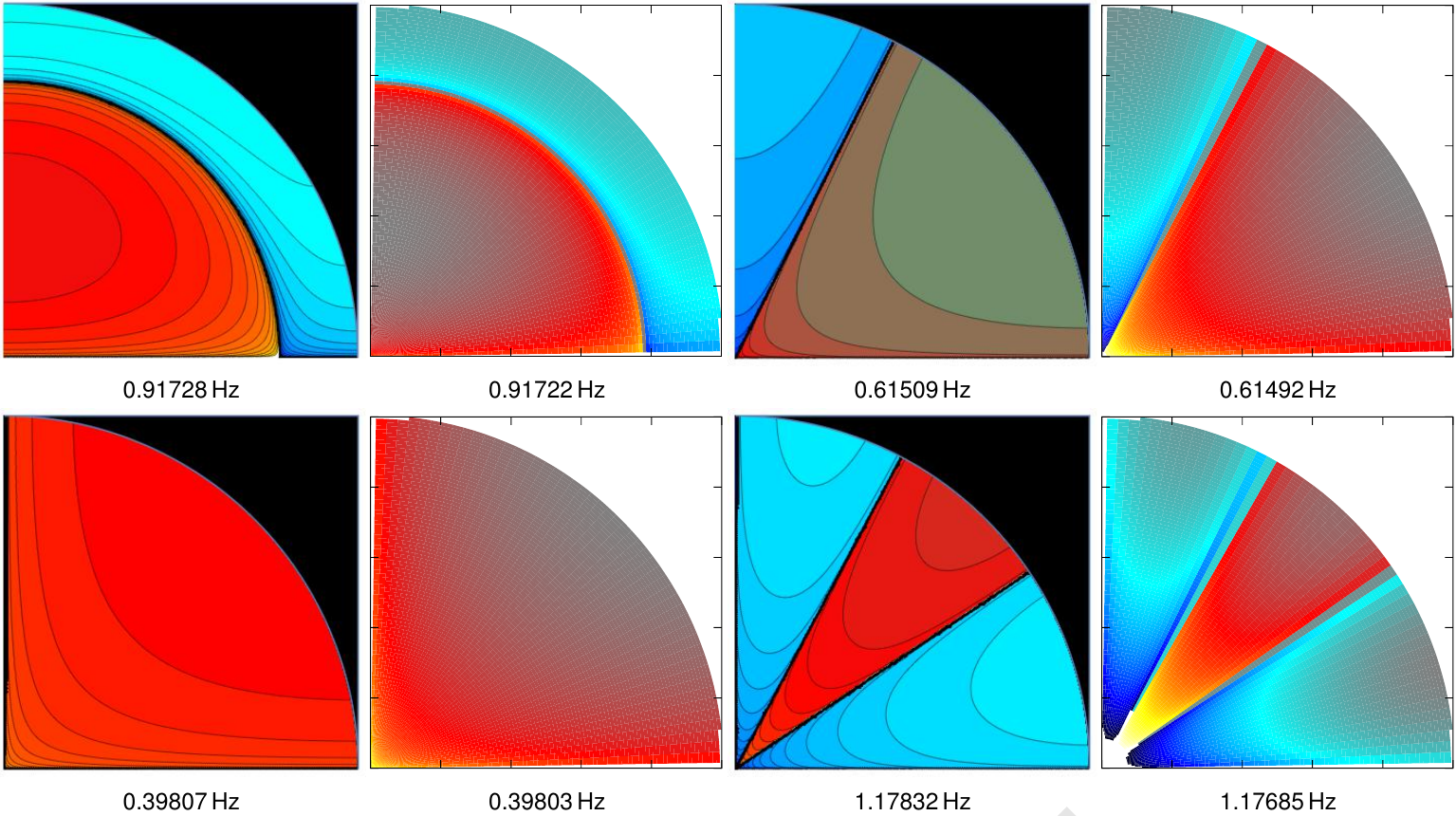}}
\caption{{\bf Callibration of the method.}
  Comparison between the analytic (panels with black background) and
  numerical (white background) solutions of vibrational
  patterns. Because of the symmetries, only one quadrant of the full
  equatorial plane of an spherical body is shown.  Modes of odd and
  even parities are displayed in the upper and lower panels,
  respectively.  In this case, we are using 100 points in the radial
  direction and 50 in the angular one. We can also observe a good
  agreement in their corresponding frequencies (listed below each
  panel), that improves as we increase the resolution.}
\label{fig:fig02}
\end{figure*}

As shown in \Fref{fig:fig02}, we get a good agreement in the toroidal
($\varphi-$) case, both in the vibrational patterns and in their
corresponding frequencies, demonstrating the ability of the numerical
code to recover the analytic values. We point out that agreement
improves with a finer mesh encompassing the eyeball (in
\Fref{fig:fig02} we employ a relatively coarse grid of $100 \times 50$
points in the $r \times \phi$ directions).  A similar analysis has
been done for modes where the displacements of the material happen
only in the $r-$ and $\theta-$directions (spheroidal modes). The
conclusion of both calibration experiments is that our numerical
procedure to compute the eigenfrequencies of the system and their
displacements is accurate enough.

\subsubsection*{Application of the method to a typical human eye}
\label{sec:application}
 
The exact eigenfrequency values are sensitive to the imposed boundary
conditions.  We assume that the surface of the eye (either the sclera
or the cornea) is free to oscillate when suitable perturbations are
inflicted to the eyeball. These perturbations can be originated by the
muscles acting either on the outer eyeball surface or on the lens
during the accommodation (e.g., contraction of the ciliary body due to
stimulation of the autonomic nervous system).  Here, we consider a set
of “standard” eye parameters. We adopt $R = 0.0125$\,m,
$\rho=1000\,$kg\,m$^{-3}$ for the eyeball typical radius and average
density, respectively. Mean values for the corneal and scleral Poisson
ratio, $\sigma$, are in the range $0.42 - 0.47$ \cite{Uchio:1999}. We
take $\sigma=0.49$, slightly above the average to account for the
incompressible character of the vitreous humour. As the
eigenfrequencies are roughly proportional to $\sigma^{-1/2}$, their
predicted values are basically insensitive to this parameter in the
typical ranges measured for constituents of the human eye. There is a
large scattering in the values of the Young's modulus, $E$, of
different parts of the eye \cite{Hirneiss:2011}. We employ a typical
value $E=0.2985$\,MPa. The eigenfrequencies exhibit a weak dependence
with the value of the Young's modulus, $\propto E^{1/2}$. Since the
largest values reported for the Young's modulus are
$E_{\rm max}\simeq 20\,$MPa, at most a factor of a few increase in the
computed frequencies is possible.

\begin{figure}[!h]
\centerline{\includegraphics[width=\columnwidth]{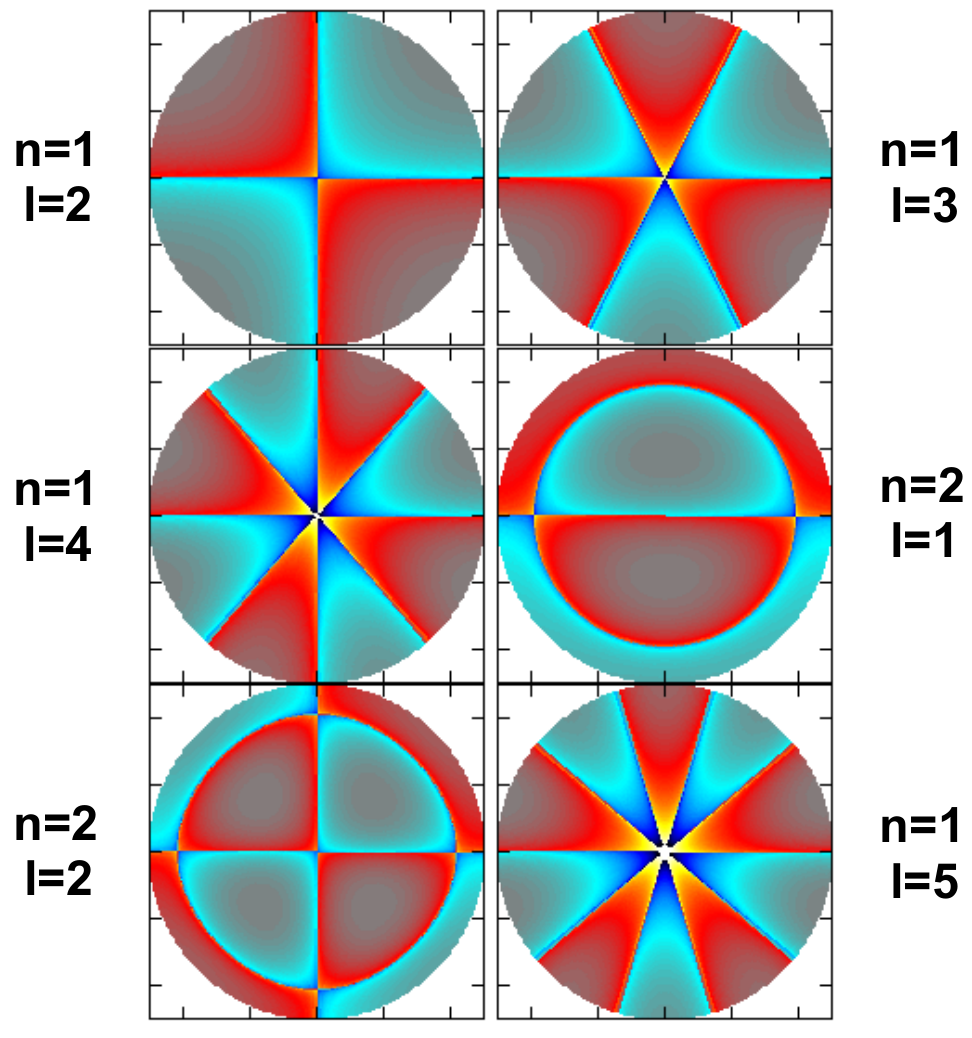}}
  \caption{{\bf Toroidal vibrational modes.} Six different patterns of
    toroidal vibration at the lowest frequencies in our model of the
    eye that correspond to the same transversal cut as shown in
    \Fref{fig:scheme}. Light and dark blue (red and yellow) shades
    indicated a motion towards (away from) the reader and normal to
    the drawn plane. {\em Left panels:} eigenfunctions with even
    parity in $l$: ($n=1$, $l=2$) vibrating at 319\,Hz, $(1,4)$ at
    648\,Hz and $(2,2)$ at 909\,Hz. {\em Right panels:} eigenfunctions
    with odd parity: $(1,3)$ at 492\,Hz, $(2,1)$ at 1159\,Hz and
    $(1,5)$ at 798\,Hz.}
\label{fig:toroidalmodes}
\end{figure}

In \Fref{fig:toroidalmodes} we show six different patterns of toroidal
vibrational modes at the lowest frequencies in our simplified model of
the eye that correspond to the same transversal cut as shown in
\Fref{fig:scheme} (for a similar figure but considering spheroidal
modes, see \nameref{fig:spheroidalmodes}). The different patterns
are identified by a set of two integer numbers $n$ and $l$ that denote
the number of nodes the solution has in the radial and in the
$\theta-$angular direction, respectively. Each pair of values $(n,l)$
has a unique characteristic frequency. The upper left panel of
\Fref{fig:toroidalmodes} corresponds to matter rotating
(counter-rotating) about the symmetry axis in the northern (southern)
hemisphere (see \nameref{fig:scheme2} for a three-dimensional
representation of the mode $(1,2)$). There is a number of normal mode
frequencies falling in the range $\sim 100\,$Hz to $\sim 10\,$MHz
(Tab.\,\ref{tab:modes}). Modes with frequencies of a few hundreds of
Hz have periods of oscillation much shorter than other quasi-periodic
variations of the eyeball volume triggered by phasic processes like
respiration and pulse.

% Place tables after the first paragraph in which they are cited.
\begin{table}[!ht]
%\begin{adjustwidth}{-2.25in}{0in} % Comment out/remove adjustwidth environment if table fits in text column.
\centering
\caption{
{\bf Frequencies of selected normal modes of the eye.}}
\begin{tabular}{|c|c|c|c|c|c|c|c|c|c|}
\hline
\multicolumn{5}{|l|}{\bf T} & \multicolumn{5}{|l|}{\bf S}\\ \thickhline
   \multicolumn{2}{|c|}{}      & \multicolumn{3}{|c|}{$l$}  &\multicolumn{2}{|c|}{} &\multicolumn{3}{c|}{$l$}          \\ \cline{3-5}\cline{8-10}
  \multicolumn{2}{|c|}{}   &       1    &        2    &      3     &\multicolumn{2}{|c|}{} &       1     &     2    &    3  \\ \hline
\multirow{3}*{$n$} &  1 & 734.46 &  318.71& 492.48& \multirow{3}*{$n$} & 1 &  2835.9 & 5706.5 &  8569.3 \\
  &  2 & 1159.0 &	909.36& 1076.1 &   & 2 & 491.41  & 947.85 & 1364.7\\
  &  3 & 1570.3 &	1339.9& 1514.1 &   & 3 &  339.58 & 694.99 &1130.0\\ \hline
\end{tabular}
\begin{flushleft} {\em Left:} Table containing the frequencies
  (measured in Hertz) of selected toroidal modes (T) computed with our
  numerical code for an average human eyeball. The set of material
  parameters employed to obtain these values are $R = 0.0125$\,m,
  $\rho=1000\,$kg\,m$^{-3}$, $E=0.2985\,$MPa, and
  $\sigma=0.49$. Toroidal modes with $n=0$ are forbidden since they
  require driving external forces (assumed non existing in this
  model). {\em Right:} Same as the left table for spheroidal modes
  (S).
  \label{tab:modes}
\end{flushleft}
\label{table1}
%\end{adjustwidth}
\end{table}

%PLOS does not support heading levels beyond the 3rd (no 4th level headings).

\section*{Discussion}
\label{sec:discussion}

In the following, we discuss first the limitations of our current
model (Sect.\,\nameref{sec:limitations}). Then, we analyze the prospects to
measure the vibrational modes of the eye with existing technologies
that where devised for different purposes, but can be suitably adapted
(Sect.\,\nameref{sec:methods2measure}).

\subsection*{Model limitations}
\label{sec:limitations}
A more accurate modelling of the eye structure than the one presented
in the Sect.\,\nameref{sec:methodology} requires differentiating (at
least) between the eye interior (including the lens and the aqueous
humour) and its elastic boundary (the cornea and the sclera). In our
model, this can be done assigning different elastic properties to
different parts of the eye. Indeed, it is possible to assign different
elastic properties on a point-by-point basis, to account for the
heterogeneity of the various eye constituents. The results of such an
elaborated model will be published elsewhere. Here, our goal is to
outline that the analysis of the normal modes may provide useful
mechanical information of the eyeball.  If we could measure variations
in the eyeball structure and if they could be attributed to normal
modes, it would be possible to set an inversion problem
\cite{Gough:1991} to obtain, for instance, the elastic moduli of the
eye. The accuracy of the solutions obtained by the inversion problem
sensitively depends on the number of properly identified eigenmodes
and on the degree of realism in the model of the eyeball. As working
hypothesis we assume that variations in the intraocular pressure (IOP)
can be used as tracers of the eyeball volumetric changes induced by
(spheroidal) normal modes of the eye.  A lot of work has been done to
connect the dynamics of the intraocular fluid by specifically
modelling the aqueous humour as a hydrodynamic system where the
inflow/outflow balance of such humour sets its physical properties,
including the IOP \cite{Lyubimov:2007}.  The variations of the
intraocular blood volume can be produced by many factors, the foremost
being pulse, respiration, IOP fluctuations, and nervous
mechanisms. The arteries of the eye are thick-walled and relatively
inelastic; thus the influence of pulse pressure on intraocular
pressure is heavily damped \cite{Maurice:1958}. Contrarily, the venous
system is thin-walled and easily collapsible and hence, its volume can
sensitively change, though in a tiny amount compared to the full
eyeball volume \cite{Weinbaum1965}.  We also point out that other
works have attempted to model only the vitreous humour as a
viscoelastic fluid, considering the vitreous chamber as a sphere, and
assuming only the effect of toroidal modes (see
\cite{Meskauskas_Repetto_Siggers_2011} and references
therein). Different from these works, we also compute possible radial
modes and present a general method that can be addapted to arbitrary
geometries.

Our model needs to be ultimately calibrated with the acquisition of
actual data of the eyeball. The ability to measure the changes in the
eyeball shape resulting from spheroidal normal modes by mechanical
means strongly relies on the maximum amplitude of the deformations
induced. In practice, the amplitude of the modes will depend on the
amplitude of the perturbations applied to the eye. As we will show in
the next section, devices developed for the continuous monitoring of
the IOP variations in glaucoma treatment \cite{Mansouri:2014} can be
used to measure the temporal variations of the eyeball volume (and,
thus, its normal modes). Since the inner eye constituents are nearly
incompressible, the spherical elastic outer shell comprised of the
sclera and the cornea must stretch to accommodate their respective
volume changes. The volume of the eyeball may change as a result of
the finite compressibility of intraocular tissue (iris, lens, ciliary
body) or due to intraocular muscular contraction. However, these
latter effects are secondary, and it is primarily the distension of
the wall of the eyeball by its incompressible contents that governs
the IOP \cite{Weinbaum1965}.

\subsection*{Methods to measure the eigenfrequencies of the eye}
\label{sec:methods2measure}

We work under the hypothesis that devices currently employed for the
continuous measurement of the IOP or suitable upgrades thereof could
be used to measure the eyeball spheroidal normal modes.  The idea of
measuring the elastic properties of the human eye taking advantage of
its internal motions has been treated from different perspectives in
the literature. The scattering pattern of attenuated laser sources on
the retina has allowed measuring the motion of the vitreous humour
\cite{Zimmerman:1980}. The shear elastic modulus could be determined
from these observations \cite{Zimmerman:1980}. Hence, the same
technique can be used to measure toroidal normal modes. The topic has
recently gained momentum since the knowledge of the mechanical
properties of the vitreous humour is instrumental for finding
materials that can be used as vitreous substitutes
\cite{Bonfiglio:2015}.  Bonfiglio et al. \cite{Bonfiglio:2015} find
resonances between the forcing frequency of their device and the
artificial vitreous.
%The sphere is made to rotate and counter-rotate periodically about a fixed axis at a given rotational frequency.  
Remarkably, high frequency resonances may result in undesirably large
values of the stress acting on the retina yielding retinal detachments
in extreme cases. These resonances are similar to the toroidal
eigenmodes we consider here. The frequencies of some of the resonances
are above 100\,Hz, in line with our results.

Toroidal normal modes of the eye are incompressible and, hence, do not
leave a trace on the IOP. The non-invasive character of swept-source
optical biometers (SSOBs) is the utmost advantage over alternative
techniques for biometric data acquisition \cite{Srivannaboon:2015}. We
foresee that the technical capabilities of SSOBs can be improved to
obtain high frequency data acquisition of the size of distinct eye
structures. Then, they could be used to identify the displacements of
the internal constituents of the eye and, therefore, to try to set an
inversion problem to recover their normal mode toroidal
eigenfrequencies.

Pulse and respiration are periodic phenomena. Therefore, they neither
affect the mean IOP nor the eyeball average volume.  The typical
frequencies of pulse and respiration are below 2\,Hz and, thereby,
they yield quasi-periodic displacements of the vascular system which
can be distinguished from the computed eyeball eigenfrequencies
(typically above 100\,Hz).  Furthermore, in order to trigger the
normal eyeball modes, it is optimal to employ perturbations having
frequencies as close as possible to the eigenfrequencies. The
perturbations induced by pulse and respiration may fall short for this
purpose if the (non-linear) mode coupling is weak.
%Nevertheless, it is theoretically possible to stimulate higher frequencies employing suitable external devices.
Micro saccadic motions of the eye can potentially trigger normal
modes, since they happen at frequencies of up to $\sim60\,$Hz,
typically last $20 - 200$\,ms and their rotational peak speeds can be
as large as 1000\,deg/sec \cite{Martinez-Conde:2009}. Micro saccades
follow the saccadic main sequence, suggesting a common generator for
micro saccades and saccades \cite{Martinez-Conde:2009}. Micro saccadic
motions are triggered by oscillatory motions of suitable frequencies
\cite{Tian:2016}.
%It is very likely that micro saccades reflect active oculomotor correction of foveal motor error, rather than presumed
%oscillatory covert attentional processes\cite{Tian:2016}. 
%
 The lowest frequencies of the normal modes of our model are close to
 the observed micro saccadic frequencies, or even closer to measured
 tremors, which consist of very fast ($\sim 90\,$Hz), extremely small
 oscillation (about the diameter of a foveal cone) 
superimposed on drifts \cite{Martinez-Conde:2009}.

There are intraocular sensors that require surgical implantation,
e.g., telemetric pressure transducer systems \cite{Downs:2011}, which
have an acquisition rate of $\sim 500\,$Hz, which may suffice for the
purpose of measuring the lowest frequency spheroidal modes in a human
eye. So far, the high-frequency IOP fluctuations have been attributed
to measurement noise \cite{Downs:2011}. Yet, our results suggest the
possibility that they are (partly) produced by the eyeball
eigenfrequencies.

Among the least invasive devices to monitor the IOP soft contact lens
sensors \cite{Leonardi:2009} (CLS) seem to be promising for our
purposes. The CLS measure at rates of 10\,Hz. This acquisition rate is
insufficient to detect the volume variations induced by spheroidal
modes. Likely, faster measurement rates are technically
plausible. However, such ability is possibly not employed because
there was no reason to provide a finer coverage of the IOP variations
so far.  Should it be technically viable to improve the data
acquisition rate in CLS based devices, they could be used for the
purpose of measuring the eye’s normal mode eigenfrequencies.

Since toroidal normal modes are incompressible, they may not be
registered by the continuous monitoring of the IOP. However, toroidal
normal modes are potentially accessible by alternative devices. The
IOLMaster 700 SSOB has shown an excellent performance
\cite{Kunert:2016,Akman:2015,Ferrer-Blasco:2016}. Should they take
measurements at high enough rate, SSOBs could be used to identify
toroidal normal modes. As the later modes yield axial displacements
about the eyeball axis, they may produce (tiny) variations of the
light propagating in a moving medium \cite{Leonhardt:1999}.  Light
rays propagating in a whirling fluid remain straight. The travel times
of rays that propagate with or against the flow differ by a
characteristic number.  The light rays differ by a certain
phase. Consequently, light waves that move with or against the medium
will show a distinct interference pattern in analogy to the
Aharonov-Bohm effect of electrically charged matter waves
\cite{Aharonov:1959}.  Perhaps it will be possible in the near future
to employ this effect in optical biometers permitting the measurement
of internal displacements of the inner constituents of the eye.

\section*{Conclusion}
\label{sec:conclusion}

We have presented the first analysis of the normal modes of an
idealized human eye to our best knowledge. For that we have imported
the analytical results developed in a number of areas of Physics, more
precisely in the field of Gravitational Wave Physics.

Finally, we will show that beyond the mechanical characterization of
the eyeball components, the normal vibrational modes of the eye could
be involved in physiological processes like, e.g., the accommodation.

The normal vibrational modes of the eye could be involved in other
physiological processes; e.g., the accommodation. Accommodation occurs
through changes in the shape and thickness of the crystalline lens.
The thickness and the curvature of the lens increase, causing an
increase in the eye's optical power.  Since it is a muscle-induced
activity, accommodation is a highly fluctuant and dynamic
process. These fluctuations are related to the fluctuations in ocular
aberrations, and occur with corresponding frequencies
\cite{Campbell:1959,Charman:1988,Dubra:2004}.  The microfluctuations
of accommodation play an important role in the variability of the
optical quality of the eye.  There are two main components of the
accommodation response: a low frequency component ($<0.5\,$Hz), which
corresponds to the drift in the accommodation response, and a peak at
higher frequency, in the $1 - 2\,$Hz band
\cite{Campbell:1959,Charman:1988}.  The vibrational eyeball modes we
have considered –having the lowest frequencies– seem to happen on
timescales of a few milliseconds.  The exact way in which the normal
eyeball modes are correlated with the accommodation process is beyond
the scope of this paper. However, we anticipate that to tackle such
study one would need to improve our current model by, at least,
differentiating in the eyeball model the cornea-sclera, the vitreous
humour and the lens. Towards this direction we will conduct our future
research.

\section*{Supporting Information}

% Include only the SI item label in the paragraph heading. Use the
% \nameref{label} command to cite SI items in the text.

\label{sec:SI}

\paragraph*{S1 Appendix.}
\label{sec:analytic}
{\bf Analytic normal modes.} In linear elasticity, the equation of motion for an homogeneous
isotropic elastic solid is given by the \emph{Navier-Cauchy} equation \cite{Love:1944}, which can be written either in vector form
\begin{equation}
(\lambda + 2 \mu) [ \nabla ( \nabla \cdot \boldsymbol{u} ) ] 
- \mu  [ \nabla \times ( \nabla \times \boldsymbol{u}) ] + \boldsymbol{F} = \rho \ddot{\boldsymbol{u}},
\end{equation}
or, component-wise, as:
\begin{equation}
\mu \nabla^2 u_i + (\lambda + \mu) \vartheta_{,i} + F_i = \rho \ddot{u}_i  ,
\label{eq:NavCau}
\end{equation}
where $u_i$ are the \emph{displacements} with respect to an
equilibrium position, $\vartheta := \nabla \cdot \boldsymbol{u}$ is
the \emph{dilatation}, double dotted quantities denote the second time
derivatives ($\partial_{tt}^2$) of such quantities, $\nabla^2$ is the
Laplacian operator, $F_i$ denote the body forces, and $\mu$ and
$\lambda$ are the \emph{Lam\'e constants}.  The Lam\'e constants are
related with the Young's modulus, $\sigma$, and the Poisson ratio, $E$, by the following expressions:
\begin{eqnarray}
\sigma =\frac{\lambda}{2(\lambda+\mu)},\qquad E=\frac{\mu(3\lambda+2\mu)}{\lambda + \mu}
\end{eqnarray}

% For a solid body, we can discretize it in a finite number of zones, each contining a given mass and being elastically bound to its neighbors. The information about the elastic properties of the discretized system is encoded into the \emph{stiffness} matrix $K$. The masses of each of the zones in the system are tabulated in the \emph{mass} matrix $M$. From the second Newton's law, which would read for each mass element $F(x) = m \ddot{x}$, and a generalization of the Hooke's law, we can write
% %
% \begin{equation}
% -K x = M \ddot{x}.
% \label{eq:masSti}
% \end{equation}

% We start assuming oscillatory solutions of the form $x = x' \cos{p t}$ and plug them into Eq.\eqref{eq:masSti}, obtaining
% %
% \begin{equation}
% -K x' \cos{p t} = - M \partial_{tt}^2 ( x' \cos{p t} ) \Longrightarrow M^{-1} K x = p^2 x 
% \end{equation}

We start assuming oscillatory solutions of the form
 \begin{equation}
u_i = u'_i \cos{ (p t + \xi)}, 
\label{eq:utest}
\end{equation}
where $p$ is the angular frequency of the perturbation, $u'_i$ are
functions independent of the time $t$ and the constant $\xi$ is
independent of the coordinates $x_i$ and only modifies the phase of
the vibration. Plugging the test solutions \Eref{eq:utest} into \Eref{eq:NavCau}, one obtains
in the absence of body forces ($F_i=0$):
\begin{equation}
\mu \Delta u_i + (\lambda + \mu) \vartheta_{,i} + \rho p^2 u_i = 0,
\label{eq:osc}
\end{equation}
which can be rewritten in the form of an eigenvalue problem 
\begin{equation}
-\frac{\mu}{\rho} \Delta u_i - \frac{\lambda + \mu}{\rho} \vartheta_{,i} = p^2 u_i.
\label{eq:eigPro}
\end{equation}
where the allowed frequencies of vibration and their corresponding
displacements (i.e., distortions of the underlying homogeneous and
isotropic structure) correspond to the eigenvalues and eigenvectors of
the Navier-Cauchy equation for given boundary conditions (imposed at
the eyeball surface; \Fref{fig:scheme}). 

There are exact solutions of \Eref{eq:eigPro} for solid elastic
bodies in which simple boundary conditions are imposed. One example is
a sphere where suitable surface tractions inhibit the momentum through
its surface (traction boundary conditions).  Under the assumption of
axisymmetry and expressing their results in terms of spherical solid
harmonics $\omega_n$, $\phi_{n+1}$ and $\chi_n$, Rue \cite{Rue:1996}
obtained, based on Love \cite{Love:1944}, analytic solutions for the
shapes of these vibrations,
%
%Here we recap the basics of such solution since they will serve for the purpose of calibrating our numerical code.
%
%We begin by defining
%%
%\begin{eqnarray*}
%h^2 := \frac{p^2 \rho}{\lambda + 2 \mu}, \qquad  \kappa^2 := \frac{p^2 \rho}{\mu},
%\end{eqnarray*}
%%
%%then we have
%%
%%\begin{equation}
%%\frac{\kappa^2}{h^2} - 2 = \frac{\lambda}{\mu}
%%\end{equation}
%%
%with which we rewrite Eq.\eqref{eq:osc} as
%%
%\begin{equation}
%(\nabla^2 + \kappa^2) u_i = \bigg ( 1- \frac{\kappa^2}{h^2} \bigg ) \vartheta_{,i}
%\label{eq:ana}
%\end{equation}
%%
%%This simplification
%%allows us to calibrate our numerical code against known analytical
%%solutions in other physics disciplines (e.g.,
%%seismology\cite{Visscher:1991} or gravitational wave
%%physics\cite{Rue:1996}).
%
%The solution to Eq.\,\eqref{eq:ana} when $u_i$ takes the form given by Eq.\,\eqref{eq:utest} can be found in terms of spherical solid harmonics $\omega_n$, $\phi_{n+1}$ and $\chi_n$ ($n=1,2,\ldots$ is the degree of the harmonic; see \cite{Love:1944}) and one arrives to:
%
\begin{eqnarray}
 u_i = \sum_{n} \bigg [ &\hspace{-0.25cm}-\hspace{-0.25cm}& \frac{1}{h^2} \frac{\partial}{\partial x_i}
  \Big \{ \omega_n \psi_n(h r) \Big \} +\psi_n(\kappa r)  \Big ( \epsilon_{ijk} x_j \frac{\partial}{\partial x_k} \chi_n  + \frac{\partial}{\partial x_i} \phi_{n+1} \Big )\nonumber \\
&\hspace{-0.25cm}-\hspace{-0.25cm}& \frac{n+1}{n+2} \psi_{n+2}(\kappa r) \kappa^2 r^{2n+5} \frac{\partial}{\partial x_i} \frac{\phi_{n+1}}{r^{2n+3}}  \bigg ]
\end{eqnarray}
as well as for the vibrational frequencies of spherical bodies. Here, $\epsilon_{ijk}$ is the \emph{Levi-Civita symbol} and
\begin{equation}
\psi_n (x) := \bigg ( \frac{1}{x} \frac{\partial}{\partial x} \bigg )^{n} \:\frac{\sin x}{x},
\label{eq:psi}
\end{equation}
%
%The exact eigenfrequency values are sensitive to the imposed boundary conditions. We assume that the surface of the eye (either the sclera or the cornea) is free to oscillate when suitable perturbations are inflicted to the eyeball. These perturbations can be originated by the muscles acting either on the outer eyeball surface or on the lens during the accommodation. So that, we impose \emph{traction boundary conditions}: the sphere is maintained in equilibrium by applying surface tractions on its boundary. Since no forces act on the surface of the sphere, the boundaries are traction free and the applied surface traction vanishes at the surface of the sphere $r=R$. ***[MAA: aren't we mixing boundary conditions for the spherical case with boundary conditions for the Cartesian case?]***
%
%With these boundary conditions, the solutions finally take the form:
%%
%\begin{eqnarray*}
%\sum_n \bigg [  p_n \epsilon_{ijk} x_j \frac{\partial}{\partial x_k} \chi_n  + a_n \frac{\partial}{\partial x_i} \omega_n + b_n r^{2n+3} \frac{\partial}{\partial x_i} \frac{\omega_n}{r^{2n +1}}  % \nonumber \\
%+ c_n \frac{\partial}{\partial x_i} \phi_{n} + d_n r^{2n +3} \frac{\partial}{\partial x_i} \frac{\phi_n}{r^{2n +1}} \bigg ] = 0
%\end{eqnarray*}
%%
%being
%%
%

In axially symmetric systems, the eigenmodes can be classified in two groups: toroidal and spheroidal. 
Toroidal modes only involve motion about the symmetry axis sketched in \Fref{fig:scheme}. Toroidal modes are incompressible since they do not change the volume of the eyeball. In this first class of vibrations $\omega_n$ and $\phi_n$ vanish ($\omega_n = \phi_n = 0$), and the frequency, from which we compute the eigenvalues of the system, is given by:
\begin{equation}
p_n= 0 \quad \mbox{    with    }\quad p_n:=(n-1) \psi_n (\kappa a) + \kappa a \psi'_n(\kappa a)\label{eq:freq} 
 \end{equation}
Spheroidal modes, implying displacements of the eyeball material in both radial and/or angular directions, are compressible. In this second class of vibrations $\chi_n$ vanish ($\chi_n = 0$). The frequency equation is given by:
\begin{equation}
b_n c_n - a_n d_n = 0
\end{equation}
with:
\begin{eqnarray}
a_n &\hspace{-0.2cm}:=\hspace{-0.2cm}& \frac{1}{(2n +1) h^2} [\kappa^2 a^2 \psi_n(h a) + 2(n-1) \psi_{n-1} (h a)], \\
b_n &\hspace{-0.2cm}:=\hspace{-0.2cm}& - \frac{1}{2n +1} \Big [  \frac{\kappa^2}{h^2} \psi_{n} (h a) + \frac{2(n+2)}{h a} \psi'_{n}(h a) \Big ] ,\\
c_n &\hspace{-0.2cm}:=\hspace{-0.2cm}& \kappa^2 a^2 \psi_n (\kappa a) + 2(n-1) \psi_{n-1}(\kappa a) , \\
d_n &\hspace{-0.2cm}:=\hspace{-0.2cm}& \kappa^2 \frac{n}{n+1} \Big [  \psi_n (\kappa a) + \frac{2(n+2)}{\kappa a} \psi'_{n}(\kappa a)  \Big ].
\end{eqnarray}
\begin{figure}[h]
 \centerline{\includegraphics[width=0.65\columnwidth]{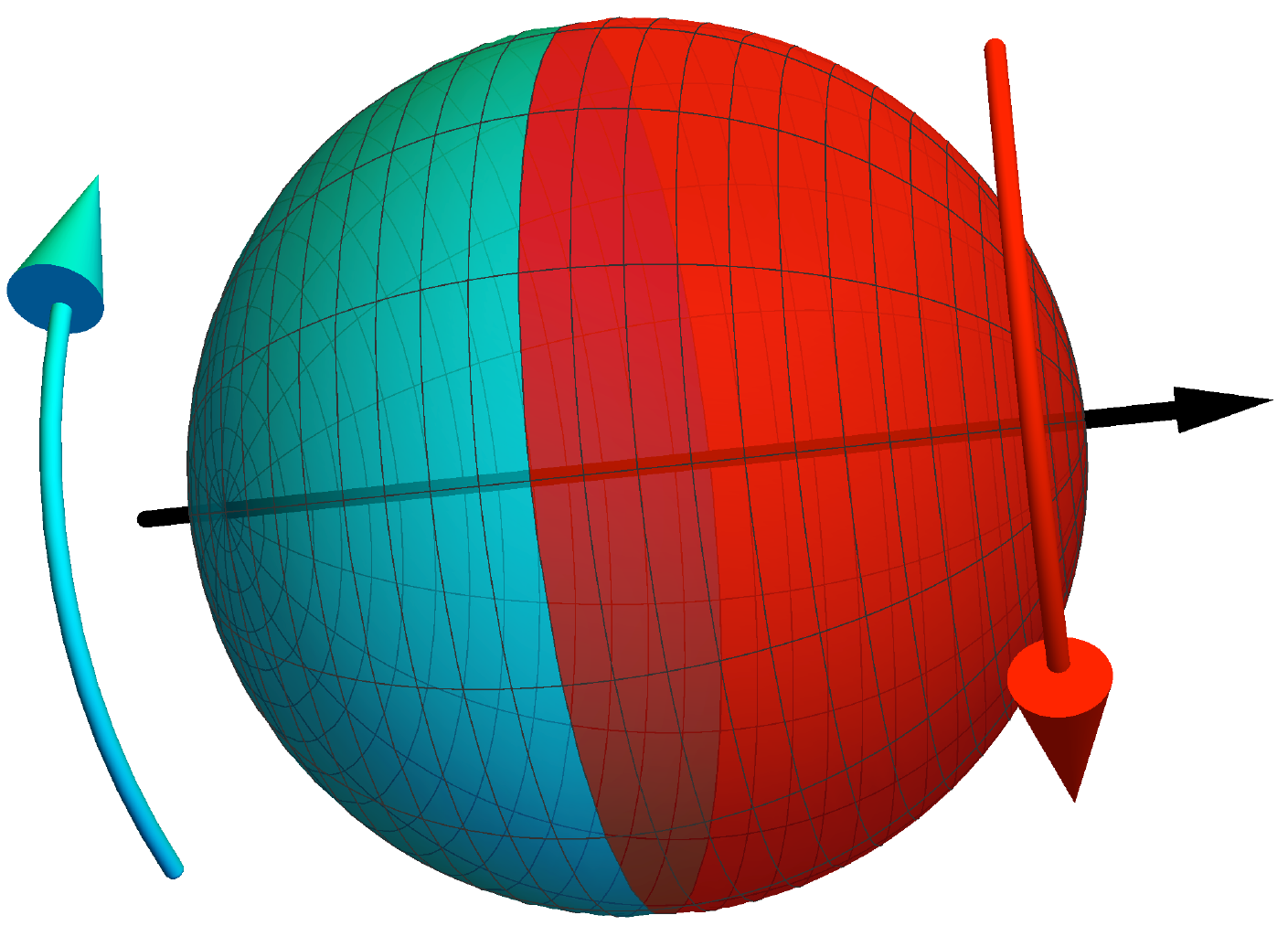}}
\end{figure}
\paragraph*{S1 Fig.}
\label{fig:scheme2}
{\bf Three-dimensional representation of the toroidal mode
     $n=1$ and $l=2$.} The mode displayed corresponds to the upper
   left panel of Fig.\,\ref{fig:toroidalmodes}. The arrows indicate
   the direction of the motion about the symmetry axis of the system
   (showed with a black arrow).

\section*{Acknowledgments}
Authors acknowledge financial support from the Spanish Government Grant Explora (SAF2013-49284-EXP).

\newpage
\begin{figure*}[h]
 \centerline{\includegraphics[width=0.85\textwidth]{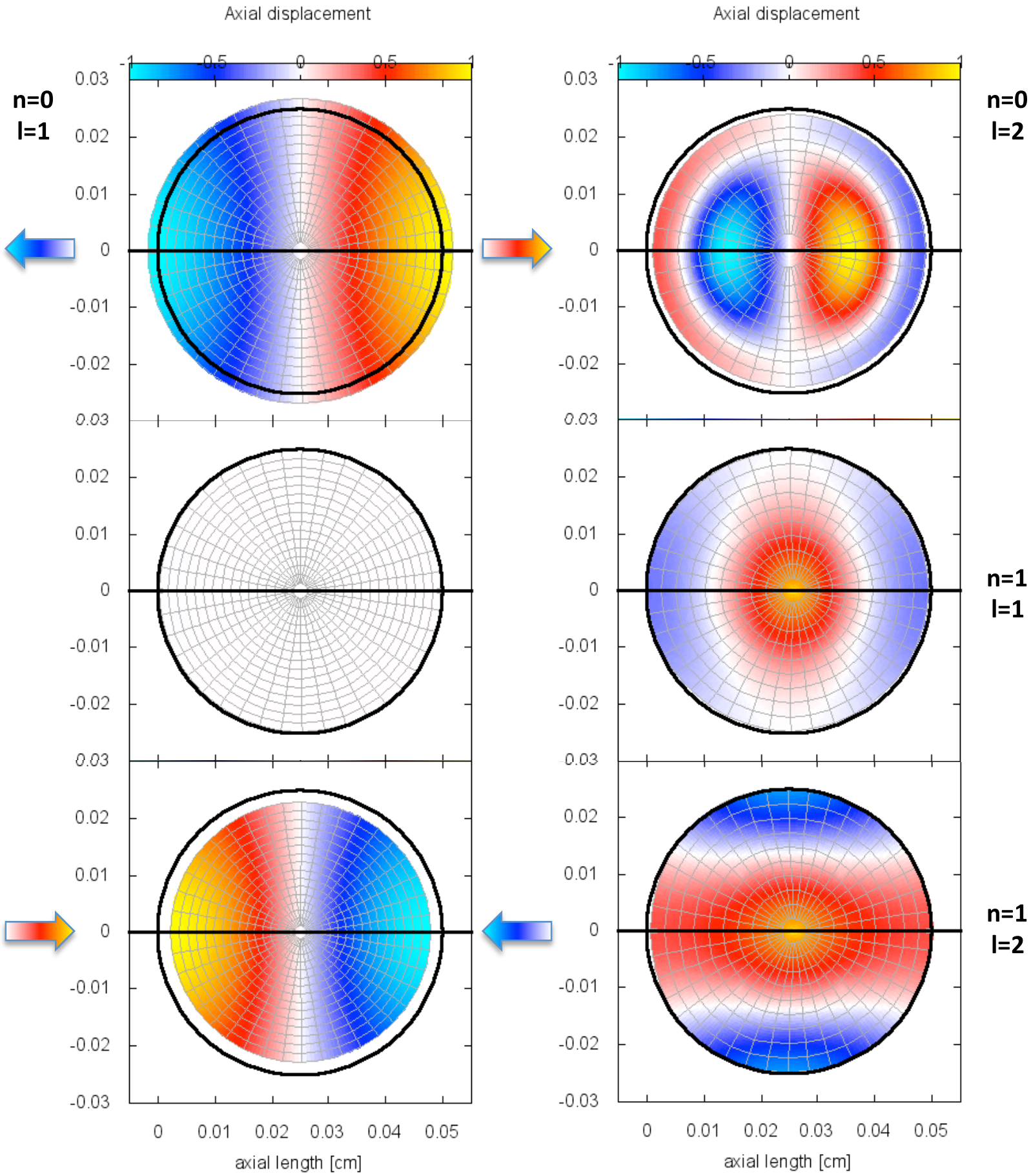}}
\end{figure*}
\paragraph*{S2 Fig.}
 \label{fig:spheroidalmodes}
{\bf Spheroidal modes.} The number of radial (angular)
    nodes is annotated by $n$ ($l$). {\em Left panels}: The eyeball
    spheroidal mode $(0,1)$, corresponding to a purely radial mode
    vibrating at $2836\,$Hz, in three different moments of its
    oscillatory vibrational pattern encompassing a half displacement
    period. We illustrate a typical vibrational period, from maximum
    expansion (top left) to maximum compression (bottom left) along
    the horizontal axis. On the central panel the displacements
    everywhere in the eyeball are null. The bottom and top panels
    correspond to times of maximum radial displacement in the
    horizontal direction. The arrows mark the direction of the
    displacements. In these left panels it is possible to observe the
    radial displacement of the boundaries with respect to the
    equilibrium state.  The maximum displacement of the eyeball
    boundary is $\sim 0.15\,$mm for the mode $(0,1)$, but this value
    is fixed for illustration purposes, since the displacement
    corresponding to a given normal mode frequency is an eigenfunction
    of the Navier-Cauchy operator, thus it possesses an arbitrary
    normalization. The quantification of the maximum radial
    displacements must be done measuring experimentally the variations
    of the eyeball shape. {\em Right panels}: Snapshots of different
    vibrational, spheroidal modes when the displacements are
    maximal. From top to bottom, we display the modes $(n, l)=(0, 2)$,
    $(1,1)$ and $(1,2)$ oscillating at frequencies $5707\,$Hz,
    $491\,$Hz and $948\,$Hz, respectively. Black circumferences mark
    the location of the eyeball boundary in the relaxed state.

\nolinenumbers

% Either type in your references using
% \begin{thebibliography}{}
% Text
% \end{thebibliography}
%
% or
%
% Compile your BiBTeX database using our plos2015.bst
% style file and paste the contents of your .bbl file
% here.
% 
\bibliography{scibib}

\end{document}